# From Neural Sensing to Stimulation: An Interdisciplinary Roadmap for Neurotechnology

Ruben Ruiz-Mateos Serrano[1†], Joe G Troughton[2†], Nima Mirkhani[3], Natalia Martínez[4], Massimo Mariello[2], Jordan Tsigarides[5], Simon Williamson[6], Juan Sapriza, Ioana Susnoschi Luca[7], Antonio Dominguez-Alfaro[8], Estelle Cuttaz[9], Nicole Thompson[10], Sydney Swedick[1, 11], Latifah Almulla[2], Amparo Güemes[1*]

[1]Department of Engineering, University of Cambridge, Cambridge CB2 1PZ, United Kingdom

[2]Institute of Biomedical Engineering, Engineering Science Department, University of Oxford, Oxford OX3 7DQ, UK

[3]Department of Clinical Neurosciences, University of Oxford, Oxford OX1 3TH, United Kingdom

[4]Department of Electrical and Electronic Engineering, Imperial College London, London SW7 2AZ, United Kingdom

[5]Norwich Medical School, University of East Anglia, Norwich, United Kingdom

[6]Department of Brain Sciences, Imperial College London, London W12 0BZ, United Kingdom

[7]Department of Biomedical Engineering, University of Glasgow, Glasgow, United Kingdom

[8]Instituto de Microelectrónica de Sevilla, IMSE-CNM (CSIC, Universidad de Sevilla), Av. Américo Vespucio 28, 41092, Sevilla, Spain

[9]Department of Bioengineering, Imperial College London, London SW7 2AZ, United Kingdom

[10]Department of Medical Physics and Biomedical Engineering, University College London, London WC1E 6BT, United Kingdom

[11]Department of Clinical Neurosciences, University of Cambridge, Cambridge Biomedical Campus, Cambridge, United Kingdom.

† These authors contributed equally to this work.
Corresponding author: Amparo Güemes (ag2239@cam.ac.uk)

**Abstract**

Neurotechnologies are transforming how we measure, interpret, and modulate brain–body interactions, integrating real-time sensing, computation, and stimulation to enable precise physiological control. They hold transformative potential across clinical and non-clinical domains, from treating disorders to enhancing cognition and performance. Realizing this potential requires navigating complex, interdisciplinary challenges spanning neuroscience, materials science, device engineering, signal processing, computational modelling, and regulatory and ethical frameworks. This Perspective presents a strategic roadmap for neurotechnology development, created by early-career researchers, highlighting their role at the intersection of disciplines and their capacity to bridge traditional silos. We identify five cross-cutting trade-offs that constrain progress across functionality, scalability, adaptability, and translatability, and illustrate how technical domains influence their resolution. Rather than a domain-specific review, we focus on shared challenges and strategic opportunities that transcend disciplines. We propose a unified framework for collaborative innovation and education, highlight ethical and regulatory priorities, and outline a timeline for overcoming key bottlenecks. By aligning technical development with translational and societal needs, this roadmap aims to accelerate equitable, effective, and future-ready adaptive neurotechnologies, guiding coordinated efforts across the global research and innovation community.

# 1. Preparing for the evolving landscape of neurotechnology

Neurotechnologies are reshaping how we understand and influence human physiology, offering tools to record, interpret, and modulate brain–body interactions for therapeutic and augmentative purposes. These systems have already demonstrated the potential to transform healthcare and wellbeing by enabling precise, personalized interventions for neurological, inflammatory and metabolic disorders, cognitive enhancement, and motor rehabilitation. Adaptive and feedback-driven systems (a.k.a. closed-loop systems) illustrate this potential, seamlessly integrating real-time sensing, computation, and stimulation to modulate neural activity dynamically. By continuously refining stimulation based on physiological feedback, adaptive neurotechnologies surpass traditional open-loop approaches, offering improved therapeutic outcomes with reduced side effects and greater personalization[1]. This includes a growing emphasis on non-invasive systems, which are increasingly sought after for their accessibility and reduced risk[2].

In clinical medicine, closed-loop systems are an established paradigm. Implanted devices (e.g. pacemakers[3] and insulin pumps[4]) are, in some specialties, standardized and effective means of restoring function to physiological systems. Yet, adaptive systems for disorders of the nervous system are still in their infancy. The neuromodulatory feedback-driven systems closest to widespread use in clinical practice are responsive neurostimulation (RNS) for epilepsy[5,6], adaptive deep brain stimulation (DBS) for movement disorders like Parkinson's disease and essential tremor and dystonia[7,8], and evoked compound action potential (ECAP) controlled spinal cord stimulation (SCS) for chronic pain[9,10]. All are iterations on existing open-loop treatments primarily aiming to match the current standard of care while minimizing side-effects or the frequency of battery-replacement procedures[11]. Alongside growing meta-analytic evidence for their use[12,13], large-scale clinical trials, are underway to support broader adoption[10,14]. Other modalities, such as Non-Invasive Brain Stimulation (NIBS), Virtual Reality (VR), Dorsal Root Ganglion Stimulation (DRGS), Sacral Neuromodulation (SNM), and Brain-Computer Interfaces (BCIs) are being actively developed with a view to converge on adaptive systems[15–18]. Increasingly, clinical trials of these technologies aim to identify sub-groups likely to respond to treatment[19], in recognition of the significant heterogeneity among neurological and psychiatric disorders[20].

At the same time, neurotechnologies are expanding beyond therapeutic use. Enhancement of attention[21], motor skill acquisition[22], learning[21], and creativity[24] resulting from neuromodulation, particularly in healthy individuals, has prompted suggestions that adaptive systems may be used for a variety of non-clinical applications across several sectors, including education, gaming, athletics, and the military[25]. This expansion has catalyzed rapid industrial growth[26]. Several companies have focused on direct-to-consumer responsive neurofeedback systems, offering devices for enhanced cognition with ergonomic form factors[27–29]. Others are developer-focused, offering software development kits, application programming interfaces, and hardware for novel adaptive applications[30,31]. A final group is focused on the development of implantable BCIs, leveraging front-end developments in microelectrode arrays and delivery routes[32–35]. This diversification illustrates the field's fragmentation but also its opportunity space. In 2024, neurotechnology companies received $1.2B in venture capital funding[36], more than any other type of medical device.

Despite this momentum, the translation of neurotechnologies from research to real-world applications exemplifies the unique challenges. A limited understanding of neural mechanisms in both health and disease constrains identification of effective targets and optimization of stimulation parameters. Added to this are methodological heterogeneity, small sample sizes, and high inter-individual variability, all of which hinder reproducibility and slow discovery. Critically, advancing the field requires integration across neuroscience, materials science, neural interfaces, electronics, signal processing, computational modelling, and clinical and non-clinical translation[37], underscoring the need for interdisciplinary collaboration.

Researchers and innovators at all career stages are essential to progress, with early-career researchers (ECRs) being particularly well positioned to drive innovation. Their interdisciplinary training enables them to bridge disciplinary silos and contribute novel solutions at the interface of physiology and neuroscience, engineering, and translation. However, the complexity of modern neurotechnology systems, from non-adaptive to feedback-driven platforms, emphasizes the need for structured, community-wide efforts to overcome barriers and accelerate progress.

This paper emerges from an interdisciplinary workshop held in Cambridge in November 2024 that convened 20 ECRs from various specializations to examine the current landscape of neurotechnologies and shape its future. We first provide a critical assessment of state-of-the-art limitations within each domain, identifying technological and conceptual barriers that constrain progress in clinical and non-clinical applications. While an exhaustive review of each individual field is beyond the scope of this work, we aim to highlight the overarching landscape and core requirements, with particular emphasis on the interplay between adaptive and non-adaptive systems. For in-depth discussions, we refer readers to specialized reviews within each area. Subsequently, we highlight key interdisciplinary bottlenecks uncovered through cross-disciplinary discussions, revealing shared trade-offs that transcend individual fields. Drawing from these discussions, we outline strategic directions for advancing the field, emphasizing the need for collaborative frameworks, standardized methodologies and protocols, interdisciplinary education and ethical considerations in neurotechnology development.

Ultimately, this paper issues a call to action for the broader neurotechnology community, across all career stages, to engage in collective problem-solving and translational research efforts. Only by aligning technical innovation with societal and clinical needs, the field can accelerate breakthroughs and shape the future of neurotechnology for both fundamental science and societal impact.

## 2. Current state of the art: overview, requirements and limitations

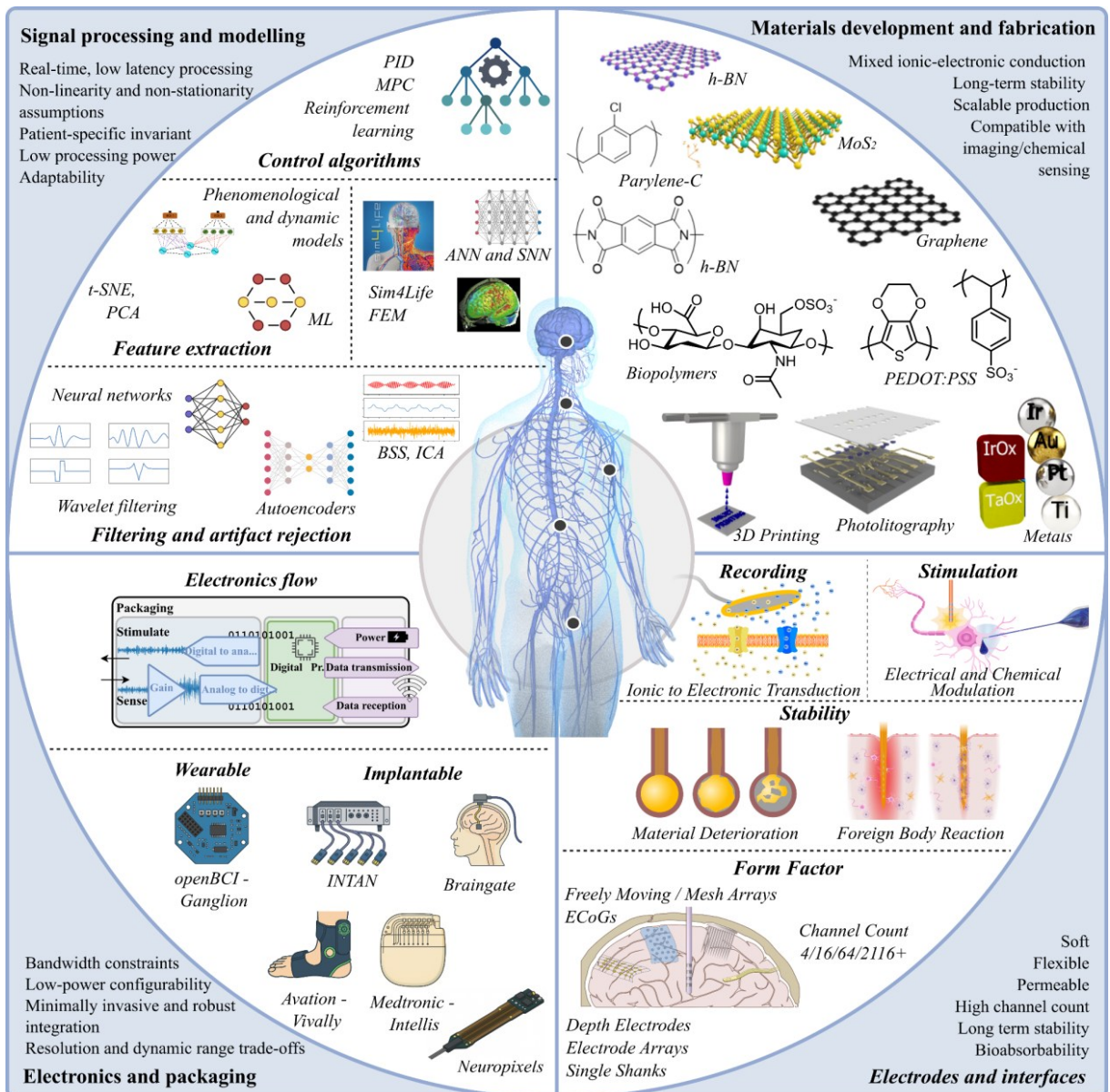

**Figure 1. Multidisciplinary landscape and technical requirements for neurotechnologies.** This diagram illustrates the current state of the art and critical design considerations across key domains required to advance neurotechnology systems: material development and fabrication, electrode and front-end interfaces, back-end electronics and packaging, and signal processing, with modelling clinical and non-clinical applications shown at the base, emphasizing current translation paths and emerging use cases such as neuroadaptive consumer technologies. The human-figure at the middle showcases current locations where these neurotechnologies are being employed, or to be deployed in the near future.

**Material development**

Neurotechnologies rely heavily on materials at the tissue-device interface (Figure 1). Traditionally, metals like gold (Au), titanium (Ti), stainless steel, silver (Ag), tantalum (Ta), platinum (Pt) alloys, and cobalt-chromium (Co-Cr) alloys have dominated due to their corrosion resistance and electrochemical properties. These materials have been approved or cleared by the U.S. Food and Drug Administration (FDA) for several neural devices treating conditions such as epilepsy, depression, chronic pain, and pelvic neuropathies[38,39]. However, as devices shrink and demand higher current densities and mechanical compatibility, metals face limitations, including increased impedance and reduced charge injection capacity when miniaturized[40]. Furthermore, stimulation can lead to toxic byproducts due to oxidation-reduction reactions and metal solubilization[41].

To overcome these challenges, advanced materials like conducting polymers (e.g., PEDOT:PSS[42]) and two-dimensional (2D) materials (e.g., graphene, MoS$_2$, and h-BN[43]) offer improved biocompatibility, electrical performance, and compatibility with nanoscale fabrication[44]. Yet, technical barriers and failure points remain for many of these materials[45]. For instance, they degrade or lose functionality when subjected to the rigours of clinical sterilization protocols or long-term implantation[46,47]. Electrochemical stability and mechanical integrity[47–49] while maintaining high performance, remains a critical challenge, as does enhancing charge injection capacity and minimizing stimulation artifacts[50–53].

Imaging compatibility is another essential consideration. Materials must not interfere with magnetic resonance imaging (MRI) and computed tomography (CT) scans, which are vital for accurate device placement, diagnostics, and postoperative monitoring[54]. While neuromodulation paired with imaging enhances precision and feedback[54,55], traditional metallic implants often cause artifacts, heating, or device malfunction[56]. This poses risks not only for clinical use but also for consumer-grade neurotechnology, such as wearables that require occasional medical evaluation. Thus, new material solutions must prioritize performance, data quality and safety across diverse applications[57–61].

**Electrodes and interfaces (front-end)**
Closely tied to materials innovation, the design and implementation of electrodes and neural interfaces remain central to neurotechnology system performance (Figure 1). Yet, limitations persist at the biotic–abiotic interface, beyond the materials choice. Early neuromodulation devices relied on labor-intensive manual fabrication methods, which remain prevalent in many commercial systems. While photolithographic techniques[55] have improved throughput and resolution, their commercial adoption is still limited due to cost and complexity, and limits in the application of emerging materials.

Electrode form factors have evolved significantly since the introduction of the Utah array in 1998[56]; however, challenges remain in balancing density, flexibility, and biocompatibility. Increasing the density of recording sites improves resolution but exacerbates mechanical mismatch with neural tissue, leading to chronic inflammation and fibrotic encapsulation. This natural inflammatory and healing response to implants is known as foreign body response (FBR) and is a major problem that limits the device functionality. Efforts to minimize implantation trauma through shape-changing devices such as expandable devices[57,58] [59,60], cortical surface devices[61], serpentine and kirigami patterns[62–64], pre-straining[65–67] and self-wrapping cuffs for peripheral nerves[68–72] have been promising, but long-term stability and seamless long-term integration remain unproven in humans.

In wearable devices, Ag/AgCl wet electrodes, utilizing conductive gel, are standard for short-term, in-lab recordings due to low skin–electrode impedance and high signal fidelity. However, they require periodic maintenance as gel dehydrates. Dry electrodes, made from materials like metals and textiles, offer a gel-free alternative enabling long-term, ambulatory use, albeit with increased susceptibility to motion artifacts and higher contact impedance. Flexible dry electrodes, incorporating foam substrates or textile-based designs, offer conformity and contact stability, especially over irregular or hairy surfaces such as the scalp. For prolonged monitoring on non-hairy regions, adhesive hydrogel-based electrodes provide stable adhesion and low impedance.[73]

**Electronics and packaging (back-end)**

Building on the foundation of materials and interfaces, back-end electronics are the bridge between the neural interface and computational systems, enabling signal acquisition, conditioning, stimulation, and data transmission. Most commercially available neurotechnologies utilize standard microcontroller-based, printed circuit board (PCB)-mounted electronics housed in large form factors[74], with part or all of the subsystem located externally, either as wearable systems[75,76] or wired to implanted electrodes[77–79]. Fully implantable systems typically require bulky units mounted in the chest or skull[80,81], connected to electrodes via long wires. These large form factors are primarily driven by the need to accommodate large batteries for power and data management, as wireless power delivery remains constrained by tissue energy absorption[82]. This challenge is further amplified by the growing demand for higher channel counts, enhanced spatial resolution, and maintaining quality of service for high-speed, accurate adaptive systems[83]. Unfortunately, the leads connected to the electrodes can introduce artifacts and signal degradation, and their integration with bulky designs results in highly invasive devices requiring lengthy surgeries and limiting scalability. Wearable systems, while less invasive and easier to deploy, offer much lower spatial resolution and specificity for recording and stimulating, are unable to access deep neural structures, and interact extensively with the external environment. Hybrid systems combining implanted electrodes and external electronics may achieve high spatial resolution[84] but remain vulnerable to environmental factors, infection risks, and limited long-term stability.

**Signal processing and modelling**

Neurotechnologies, and in particular closed-loop systems, rely on real-time signal processing and control algorithms to extract actionable information from noisy, dynamic neural data[85,86], which remains a major challenge (Figure 1). Traditional denoising methods like wavelet transforms[87], independent component analysis (ICA)[88,89] and Empirical Mode Decomposition (EMD)[90–92] have enabled progress in applications like seizure detection but often fall short in handling the complexity and non-stationarity of neural signals, particularly under real-time constraints. Spatial filtering methods such as beamforming offer faster artifact rejection[93,94], but rely heavily on accurate source modeling[95]. Autoencoder-based models have been proposed to address this gap, but these approaches still struggle to balance artifact removal with the preservation of critical neural signal components[96]. Emerging hybrid approaches, such as integrating spiking neural networks (SNNs) with wavelets[97,98], adaptive neural networks[99] and state-space models[100], offer improved responsiveness but are often too computationally intensive for real-time, multi-channel applications. Feature selection is another hurdle: automated frameworks, like *catch22*, can extract statistical features[101], but clinically relevant biomarkers are still typically found through trial and error, as in DBS[102]. While machine learning models, including autoencoders and reinforcement learning, show promise in tailoring neurostimulation, they often require extensive validation and are limited by poor generalization across species[103]. Phenomenological computational frameworks integrating dynamical models with machine learning have shown potential in extracting robust biomarkers for real-time neurostimulation[104,105]. Control algorithms have advanced from basic Proportional-Integral-Derivative (PID)[106] to model-predictive[107] and reinforcement learning-based[108,109] systems, yet real-time application in humans remains rare due to computational load in real-time applications. A general limitation is the lack of generalization to human applications due to interspecies differences in neural dynamics[110]. Standardized validation protocols, coupled with methods like transfer learning, are therefore essential for improving translatability[111,112]. As the field matures, a key priority is the development of lightweight, adaptive algorithms that can operate in real-time, generalize across patient populations, and integrate seamlessly with sensing and stimulation hardware.

**Ethics and regulatory landscape**

The rapid advancement of neurotechnologies offers unbridled potential for treating diseases, from restoring mobility in paralysis to managing chronic pain. However, as UNESCO highlights[26], neurotechnologies challenge core aspects of human existence, including mental integrity, dignity, autonomy, and privacy. The very nature of interfacing with the nervous system and the brain, the epicenter of our very being, increases risk of harm and introduces considerable ethical and regulatory challenges. Despite a 268% rise in neurotechnology-related patents between 2013 and 2022[113], an almost 35-fold increase in neuroscience publications between 2000 and 2021[26], and a huge surge in investment from governments worldwide, ethical and regulatory processes remain slow. The evaluation of complex, dynamic systems which comprise materials, electronics, signal processing algorithms, and software poses significant challenges for existing regulatory frameworks, which remain ill-equipped to address these multidisciplinary demands.

In the UK, the Health Research Authority (HRA) ethical approvals average 74 days for non-commercial and 113 days for commercial studies[114], with subsequent regulatory approvals from agencies such as the Medicines and Healthcare products Regulatory Agency (MHRA) often exceeding a year for Class III implantable devices. In the European Union (EU), the Medical Device Regulation (MDR) provides the legal foundation for regulating medical devices including neurotechnologies. In 2022, the MDR expanded its reach to also include certain non-medical devices including non-invasive brain stimulation systems that may be used for purposes such as gaming or well-being[115]. This expansion introduced performance-based safety standards and a 'maximum acceptable risk' threshold, rather than requiring therapeutic benefit. For clinical neurotechnologies, the requirement of clinical evidence, post-market surveillance, and notified body assessments, leads to delays in approval exceeding 18 months for Class III devices. In the US, the FDA regulates neurotechnologies via the premarket approval (PMA) and de novo pathways. The Breakthrough Devices Program offers an expedited pathway for transformative technologies, but approval for Class III devices still averages 12 to 18 months, with investigational device exemptions (IDEs) adding further time before clinical trials.

Data privacy, protected under GDPR (EU) and HIPAA (US), is critical given the sensitivity of neural data. Although neurotechnologies promise transformative advances, they face considerable ethical and regulatory barriers. Prolonged approval processes, high development costs, and fragmented international regulations restrict access and hinder innovation.

**3. Cross-Cutting bottlenecks and trade-offs in neurotechnology**

With the landscape and domain-specific limitations now established, it becomes evident that the most formidable barriers to progress in neurotechnology do not exist *within* these silos, but between them: interdisciplinary bottlenecks that reflect competing demands among performance, scalability, usability, and clinical viability. Unlike most medical devices, these systems interact with a highly adaptive nervous system, requiring simultaneous precision in recording and/or actuation, and continuous adjustment to patient-specific neural states. This section dissects the fundamental trade-offs that emerge at the intersection of these disciplines, which represent the true grand challenges for the field and outlines potential strategies to mitigate or resolve them.

## 3.1. Trade-off between material functionality and long term stability

Reliable long-term communication between neural tissue and electronic devices is key for adaptive and non-adaptive neurotechnology. Achieving long-term biocompatibility remains challenging, primarily due to FBR degrading signal quality over time. This challenge is compounded by the extremely small amplitude of neural signals, often in the microvolt range, and the need for devices to interface across delicate barriers such as the blood-brain barrier, which adds further constraints on material choice, size, and invasiveness. To overcome this, future neural interfaces must incorporate materials that reduce impedance and noise while maintaining flexibility and matching the mechanical properties of the surrounding tissue, thereby improving long-term stability and real-time feedback integration without frequent maintenance or replacement. Materials should also exhibit long-term stability against biofluid permeation or corrosion[116,117] to operate reliably over extended periods. Equally, for wearable BCI devices, materials must maintain exceptional long-term performance and user experience while ensuring user safety and comfort[118–120].

Materials with mixed ionic-electronic conductivity have demonstrated high quality signal transmission from the tissue to device circuits. PEDOT:PSS stands out as a soft and flexible candidate with capacity to swell in water-based media that presents mixed ionic-electronic conductivity. In scenarios where the front-end device is fully polymeric, the water permeability and the mix condition of PEDOT:PSS enhances neural signal recording[121]. However, this swelling also can compromise the stability of conventional silicon-based electronic connections at the device's back-end hardware (electronics and packaging). This problem is further aggravated by micromotion between the implanted devices and the soft neural tissue, causing inflammation, gliosis, and eventual signal loss[122,123]. Addressing these issues requires advanced encapsulation strategies that balance flexibility with hermetic sealing while minimizing adverse immune reactions[123,124]. Beyond fully polymeric implementations, PEDOT:PSS can also be employed solely as a coating for metal electrodes, such as plasma-treated gold. Although PEDOT:PSS lacks functional groups for covalent bonding to metals, the silane-based crosslinker GOPS has been demonstrated to react orthogonally with the PSS unit and hydroxyl groups on plasma-treated metal surfaces[125] to obtain an improved signal-to-noise ratio (SNR) and ensure good-quality data[126–128].

On the other hand, materials optimized for neural recording often struggle with charge injection requirements for stimulation, whereas materials engineered for neuromodulation may suffer from high impedance and diminished signal fidelity[129,130]. PEDOT:PSS presents properties that facilitate effective recording and stimulation, two core functionalities in neurotechnology systems[131]. However, when used for coating metal surfaces, the resulting chemical bond exhibits poor adhesion and instability under hydrolysis conditions, particularly under repeated electrical stimulation[49,132]. Electrical stimulation protocols also impact material longevity, as repeated charge injection can lead to electrode degradation, increased impedance, and signal instability over time[45,130,133,134]. Optimizing stimulation waveforms, charge balancing, and material composition is crucial in preventing long-term performance deterioration, particularly for chronic implants[135]. Recently, ambipolar and p-type organic semiconductors have emerged as promising alternatives to improve the efficiency and versatility under recording or stimulation, enabling the transport of electrons as the dominant charge carriers while avoiding issues with regard to mix conduction[136,137].

Bioresorbable materials constitute another promising pathway, enabling temporary therapeutic solutions without necessitating surgical removal. These materials seamlessly integrate with neural tissues and degrade harmlessly over time. Despite this main advantage, their development and deployment have been limited in clinical contexts, primarily due to difficulties in controlling degradation rates. Biodegradable polymers and bioresorbable metals must maintain mechanical integrity and functional performance for a precise duration before degrading at a predictable rate. Inconsistencies in degradation can lead to premature failure or inflammatory responses. Additionally, achieving biocompatibility while ensuring robust electrical conductivity and signal fidelity remains difficult, as many biodegradable materials possess inherently poor electrical properties, limiting their effectiveness in recording and stimulating neural activity. Ensuring controlled degradation without sacrificing performance requires meticulous material design[138,139].

In parallel, efforts towards miniaturizing neural interfaces are shifting these devices from PCB-based systems to wireless millimeter-scale ASICs with compact, hybrid, or flexible packaging. These advances enable minimally invasive implantation, improve biocompatibility by reducing FBR, and enhance long-term stability. Importantly, these developments increasingly integrate diverse materials and functionalities, such as embedded sensing and on-node processing, into a unified platform, advancing the long-term performance and versatility of closed-loop systems across both central and peripheral applications.

*Resolving this trade-off requires a paradigm shift towards co-design, where material properties and long-term stability are considered inseparable from the initial stages of device engineering.*

**3.2. Trade-off between innovative development and scalable manufacturing**

Transitioning novel technologies into clinical practice is a long journey as the new materials and device form factors being developed often rely on emerging fabrications technologies that are not yet widely accessible for scalable manufacturing. Currently, few facilities are available to scale up device production to a level where comprehensive pre-clinical testing can be carried out.

Aside from commercially available formulations like PEDOT:PSS, achieving scalability and reproducibility with novel materials remain significantly challenging. Unlike the well-established silicon-electronics manufacturing, which are based on iterative materials deposition and photolithography steps to build complex, multi-component electrical interfaces, the development of advanced materials, such as 2D materials[140], n-type conjugated polymers, and bioresorbable substrates, requires specialized, lengthy, and synthetic pathways coupled with precise manufacturing to control thickness and uniformity, all of which directly influence device performance[141,142]. However, existing microfabrication and printing techniques are not yet optimized for the seamless integration of heterogeneous materials, miniaturized geometries, and intricate form factors. The complexity of integrating bioresorbable materials with wireless power sources and adaptive systems further intensifies these manufacturing challenges.

Meeting the diverse performance demands of neural interfaces often requires combining multiple specialized materials, which adds fabrication complexity and introduces more potential points of failure. Although recent advances have benefited from silicon-based manufacturing methods, these approaches limit design flexibility and form factor. Moreover, many emerging materials for neural applications are incompatible with conventional semiconductor processes, posing additional integration challenges. Recent trends have shifted toward innovative manufacturing methods that bypass traditional dry and wet etching steps, allowing simultaneous integration of biopolymers and hydrogel-like structures with miniaturized form factors. These increasingly popular techniques for researchers include printed electronics methods such as inkjet[143] and screen printing[144], conventional extrusion-based 3D printing[145–147], two-photon polymerization, laser machining [68,148–150], and even in situ fabrication of device components. However, these tools often remain confined to specialized, expensive in-house applications. Integration with microelectronics must additionally meet requirements related to power consumption or heat generation, which are related to tuned material thicknesses, adjustable designs and form factor, as well as the assembly of conductive and dielectric materials[151].

Although individual research institutions can readily achieve small-scale demonstrations of novel devices, wider pre-clinical investigation and validation require dedicated small-volume manufacturing facilities to transition effectively from academic research to commercial products. While there are a number of contract manufacturing companies that may fulfil this role, accessing these can be challenging. Establishing dedicated scale-up facilities for device production presents a preferable alternative to contract manufacturing. The Center for Process Innovation (CPI) is a UK-based example of such a facility, aimed at a wide range of technologies including flexible electronics and healthcare technologies. While CPI can act as a model in supporting the scale up of some technologies, their capacity and tooling is limited.

Beyond scale-up requirements, an even greater shortfall exists in manufacturing capacity for clinical-trial-ready devices. While a number of global MedTech companies have capacity in-house for the production of Class II & III active medical devices, primarily for control and processing electronics, this remains inaccessible for most development purposes. To promote global development of new devices, tooling must be made available to a wide range of academic and commercial users, rather than being isolated within individual research groups. For a wider pre-clinical trials, platforms must be provided for scaling up both materials and device production, and sharing knowledge of these pipelines. Furthermore, complementary facilities must be developed, equipped with the necessary tooling and regulatory compliance frameworks, to produce trial-ready medical devices.

Economically, synthesis and development of innovative materials, implementation of new device designs, and miniaturization of neural interfaces require substantial funding and a robust economic framework. Clinical-grade facilities adhering to stringent quality control standards such as Good Manufacturing Practices (GMP) and Quality Management System (QMS) are necessary for testing and human trials, even for conventional materials like platinum. Additionally, the back-end electronics of neural interfaces necessitate advanced ASICs with high TRL levels, as well as powerful hardware to process acquired data. These factors collectively slow the translation of neuromodulation devices.

While these challenges are known to be costly, there is a lack of detailed analyses regarding the funding required to advance ideas from concept to clinical translation. Financial constraints impact not only the testing, validation, and translation of new technologies but also the scalability and integration of these devices into real-world applications. To date, only one study has analyzed the cost-effectiveness of closed-loop SCS compared to open-loop systems, concluding that adaptive systems provide more quality-adjusted life year at a lower cost[152]. Expanding such analyses to other neuromodulation applications could provide critical insights, fostering collaborative funding strategies, public-private partnerships, and efficient resource allocation. These efforts could accelerate innovation while ensuring economic feasibility. Moreover, demonstrating cost-effectiveness during clinical use could further motivate adoption and highlight the economic advantages of closed-loop systems over conventional open-loop alternatives.

*Resolving this trade-off demands coordinated investment in accessible fabrication infrastructure and scalable design pipelines that bridge academic innovation with clinical-grade production.*

**3.3. Trade-off between spatiotemporal resolution and real-time capabilities**

The effectiveness of neuromodulation, and particularly closed-loop neuromodulation, depends on how closely recorded signals reflect the underlying neural activity and how precisely stimulation is delivered to target neural circuits. Efficient acquisition, processing, transmission, and actuation of increasing amounts of data is therefore critical. Adaptive neuromodulation systems must therefore balance power size and latency requirements to achieve clinical uptake.

Increasing the number of electrodes enhances recording and stimulation precision but comes at a cost. Higher channel counts demand complex amplifying electronics, escalating power consumption and device footprint[153–155]. Multiplexing techniques reduce backend complexity but degrade signal quality in the time or frequency domains[156–158]. This trade-off between fidelity, cost, and power efficiency remains a major hurdle for real-time, high-density neural interfacing.

Ideal electrodes accurately capture neural signals while minimizing external interference. A high dynamic range allows the system to detect weak neural signals and strong transient artifacts without losing critical information. This requires advanced signal processing techniques where amplification and filtering can also introduce additional noise, ultimately degrading signal quality.

Adaptive channel selection is another key enabler, continuously activating the most informative channels while suppressing noisy or redundant data, enhancing spatial resolution. This is particularly beneficial with the proliferation in high-density electrode arrays, necessitating the extraction and processing of large quantities of neural activity. Nevertheless, the quality of the acquired data still heavily relies on electrode design and placement. Poor electrode placement or design can lead to high impedance and noise, reducing recording quality.

Data throughput can be enhanced by increasing on-chip processing, but this comes at the cost of higher power requirements. Increasing the available power, however, remains a challenge, particularly in wireless systems. Electromagnetic power transmission is the most mature method, but is limited by battery life, restricted range for temperature increases, and antenna footprint as well as by tissue absorption (in implantable devices). Proper resource management is thus crucial to enable portable, power-efficient devices that are more likely to be adopted in clinical practice. Alternative power and data transfer strategies have been proposed including high-frequency ultrasound and near-infrared, although each carries inherent trade-offs related to biocompatibility and integration complexity. Advances in predictive modulation and wireless data transmission also help to minimize data transmission requirements, conserving both energy and bandwidth.

The data requirements for closed-loop neuromodulation are often vast, exceeding the capacity of current processing architectures, particularly the need for real-time processing and feedback. While deep learning and high-dimensional signal processing approaches provide the necessary performance, their power requirements remain well beyond what is available in stand-alone devices. Current research is exploring a variety of innovative solutions to this. Hardware acceleration, particularly through GPUs and FPGAs, can provide the necessary computational power, while algorithmic optimization ensures that only the most relevant data is processed. Hybrid approaches, which dynamically adjust model complexity, offer the flexibility to balance performance and efficiency. Additionally, compression methods such as adaptive sampling and time-division multiplexing allow systems to reduce the amount of data they handle, significantly lowering power consumption. Meanwhile, machine learning techniques focused on feature extraction offer the potential to automatically identify and prioritize important data features for more efficient processing.

Finally, neuromorphic computing, inspired by the brain's architecture, is emerging as a promising avenue to reduce power requirements for on-chip processing, offering a route to energy-efficient, low-latency processing capabilities for both signal detection and neurostimulation[159].

*Resolving this trade-off requires a multi-pronged approach combining optimized electrode design, adaptive signal processing, and energy-efficient architectures, such as neuromorphic computing, to balance resolution, latency, and power consumption in neurotechnology systems.*

### 3.4. Trade-off between multiscale multimodal integration and adaptability

Advancing neurotechnologies requires greater adaptability across physical and software domains to accommodate a wide range of deployment scenarios, during both data acquisition and stimulation, while accounting for intra- and inter-user variability. Adaptability in neurotechnologies is also uniquely challenging because the brain and user are themselves adaptive across many timescales, from rapid synaptic plasticity and state-dependent modulation to long-term learning and behavioral changes. Devices therefore face a moving target, where performance must be maintained despite ongoing neural and behavioral adaptation. This inherent adaptability of the nervous system directly conflicts with scalability: with limited training data, systems are unlikely to generalize well across diverse users or scenarios, and accounting for such variability requires large and heterogeneous datasets. Effective platforms must therefore dynamically adjust to diverse temporal resolution and spatial coverage demands across multiple modalities, enabling precise targeting, accurate event detection, and personalized responses. Adaptability also extends to applications like brain-computer interfaces, well-being monitoring, and gaming. To ensure generalizability, systems must perform reliably across populations with diverse physiological traits (e.g., hair type, skin conductance, age, gender) while balancing wearability, form factor, cost and data fidelity.

However, developing more adaptive systems faces a deadlock: high-quality results typically require extensive data acquisition from hundreds of channels, which conflicts with size and power constraints. A way forward is through distributed systems, where minimally invasive, flexible, and miniaturized sensors collect data close to the zone of interest while conforming to the body's shape and movement[160]. These can support multimodal data acquisition and adaptable stimulation, enabled by low-heat dissipation electronics and wireless power and data transfer (WPDT). Processing is increasingly offloaded to the edge: local feature extraction, channel selection occurring near the sensor (on-the-fly processing), and event-based sampling minimize transmission and power. Although current signal processing techniques are not fully compatible with event-driven methods, asynchronous approaches like spike-based neural networks (SNNs) and event-driven systems[161] can bridge this gap.

More powerful computational platforms can be placed in less-constrained locations (e.g., chest units or wearable devices) ensuring low-latency inference, improved energy efficiency, and data privacy. CPU-based platforms are suited for evolving inference algorithms without hardware translation overhead, while reconfigurable circuits and neuromorphic computing enable real-time processing through parallelization, data quantization, and memory-efficient designs (e.g., SRAM over DRAM). Fully neuromorphic pipelines, from feature extraction to decision-making, allow for low-latency and energy-efficient operation while preserving algorithmic flexibility.

While recent advances in neuroprosthetic systems, such as speech decoders[162], demonstrate remarkable capabilities, most traditional approaches that rely solely on electrical recordings still face limitations in capturing the rich biochemical and physiological context underlying neural function Integration of optical and biochemical sensing provides real-time monitoring of metabolic, hemodynamic, neurotransmitter, and inflammatory markers for a greater understanding of the physiological environment and signal dynamics, which are increasingly recognized as important for understanding the physiological environment and supporting the long-term stability of adaptive implants. Combining modalities (e.g., electrical, optical, chemical) enables more comprehensive interaction with neural systems, allowing both sensing and stimulation to be tailored to the specific physiological phenomenon of interest. Different approaches bring distinct strengths and limitations. For example, imaging techniques (e.g., fMRI, calcium imaging) provide high spatial resolution but slower dynamics, while chemical sensing (e.g., neurotransmitters, pH) yields rich contextual data but may be affected by signal drift or lag. Rather than requiring every system to achieve all functions, the choice of modalities depends on the target process and intended therapeutic or scientific application. To overcome these issues, a promising approach involves combining phenomenological models of delay-coupled nonlinear systems with forward models (e.g., Balloon Windkessel). These frameworks account for dynamic interactions and time delays, enabling comparison of distinct features that are captured across multiple modalities.

Material choice also critically influences multimodal integration, as many conventional electrode metals produce imaging artifacts. Promising electrode materials, such as carbon materials[163] (e.g., graphene, carbon nanotubes) as well as conducting polymers[164] have shown strongly reduced artifacts for both MRI and CT. Alternatively, optical fibers, replacing wires for photonic power transfer have demonstrated full MRI safety[165]. Integrating multifunctional materials for multimodal sensing, stimulation, and drug delivery is also a promising avenue. Materials such as magnetic nanoparticles for magnetothermal neuromodulation and plasmonic nanoparticles for optogenetics allow real-time biochemical monitoring optimized via machine learning[166]. While hybrid designs incorporating bioresorbable, ionic-electronic, or optically active components hold great promise, they require novel encapsulation techniques, often compromising application-specific performance.

Given patient response variability, identifying structural and functional neuroimaging biomarkers is key for adaptive neuromodulation. Hybrid models, combining finite element analysis (FEA) with neural networks, offer promising solutions for integrating real-time stimulation dynamics with adaptive responses, addressing scalability and computational challenges[167]. These techniques improve the modeling of nonlinear interactions across networks and capture multi-scale dynamics, advancing adaptive applications while maintaining the physiological accuracy required for clinical and non-clinical implementations[168]. Cloud platforms may support the aggregation of multiscale data by combining macroscopic connectome data with neural population models across patients, but this approach must be carefully balanced against concerns around data privacy, autonomy, and system dependency. Rather than relying on continuous internet connectivity, we envision secure, locally hosted or edge-computing solutions that enable patient-specific calibration and algorithm training without compromising user agency or safety. Achieving effective multi-modality generalization and personalization requires large, standardized datasets and proper testing frameworks. Non-clinical applications have the power of providing data from a wider range of scenarios for this cause, enabling more streamlined personalization to future users.

*Resolving this trade-off will require distributed, event-driven architectures, multimodal sensing platforms, and hybrid modelling frameworks that together enable scalable personalization and robust adaptability across diverse physiological and deployment contexts.*

### 3.5. Trade-off between technological complexity and clinical or commercial translatability

In general, the development of the next generation of neurotechnology systems follows two pathways: developing novel neurotechnologies from in silico/in vivo preclinical models, or upgrading existing open-loop solutions to incorporate closed-loop capabilities. In either case, the effective translation of neurotechnology systems for clinical and commercial use depends fundamentally on high-quality evidence, ethical and regulatory approval[169–171], and stakeholder acceptance. The relevant stakeholders include patients, clinicians, regulatory bodies, healthcare providers, researchers, and commercial end-users such as gamers. Each group plays a crucial role in the adoption and integration of these technologies into practice. Ensuring acceptance and usability requires careful consideration at all levels of development, from material selection and device design to data interpretability and user interfacing and comfort. Technological complexity at any level can be a barrier to translation.

At the level of materials, significant complexity arises from interactions between device components and biological tissue, particularly for implanted devices, which carry the risk of an adverse tissue response; conventional rigid electronics can also restrict mobility and cause discomfort. Even minor differences in material composition can lead to inconsistencies in device performance, complicating cross-study comparisons and hampering the development of standardised manufacturing processes. The greater the complexity of these interactions and their downstream effects on device function, the greater the likelihood of additional regulatory bottlenecks and limited widespread adoption[169–171]. Material design must also be optimized for efficient and safe surgical handling during implantation procedures, which, in turn, must minimize the risk of tissue damage and be simple enough to be performed routinely. Diversity of form factors requiring bespoke procedures will result in discrepancies in clinical outcomes[172].

Once implanted, maintenance of the device is an important concern. Many current neurotechnology systems require regular recharging or invasive battery replacement. This in turn increases patient burden, limits accessibility, and raises ethical questions regarding equitable care distribution[170,173,174]. Furthermore, device failure, such as through biofouling or material degradation, not only compromises function but also requires additional invasive procedures to remove or replace the device, introducing further clinical and ethical complexities[170,175]. Without ongoing support, these devices risk becoming "abandonware" (i.e. remaining in the body without updates, maintenance, or serviceability) posing significant threats to patient safety[173,174]. User-driven abandonment is also a concern: devices considered intolerable may be disengaged before meaningful long-term data regarding their effectiveness can be gathered[176,177]. To prevent this, patients should be informed of the uncertainty regarding the long-term risks of implanted devices before providing consent, and manufacturers must ensure compliance with ISO standards as a benchmark for safety and reliability to comply with stringent regulatory frameworks[171].

Regarding data and power transmission, wireless technology has well established protocols from internet-of-things (IoT) with several explored modalities (WiFi, BLE, radio, etc.); however, a balanced solution between wireless power transfer and wide data-rate remains underdeveloped across both clinical and commercial settings[171]. Existing regulatory approval frameworks often struggle to accommodate emerging power technologies, delaying the adoption of innovative energy solutions[169,170], and causing disparities in access to safe and sustainable power sources[115,173,175].

The signal processing and adaptive algorithms necessary for closed-loop systems are additional sources of technological complexity. A direct strategy is upgrading open-loop systems by leveraging existing clinical and technical infrastructure, allowing developers to incrementally model and refine the feedback and control elements specific to adaptive operation. However, this path presents unique challenges. Retrofitting closed-loop functionality into existing platforms may be constrained by hardware limitations, lack of integrated sensing, or insufficient understanding of the appropriate biomarkers and control algorithms. As a result, upgrades often require significant redesigns to support real-time data acquisition, processing, and actuation while maintaining clinical usability and regulatory compliance.

On the other hand, for novel technologies, a major barrier is accurately modelling healthy or diseased neurophysiology. This is essential to effectively translate preclinical strategies, but challenging as neural processes are predominantly nonlinear[178] and anatomical variations can limit direct translation of devices and stimulation dosages[179,180]. AI-enhanced personalized imaging may enable detailed mapping of the brain and peripheral nerves, which is likely crucial for individual-tailored adaptive systems. With fully mapped and characterized neural substrates, it becomes possible to implement real-time feedback systems for accurate sensing, monitoring, and adaptive stimulation control. However, complex models and their associated systems present ethical and regulatory concerns regarding data privacy, algorithmic transparency and generalizability[115,173,174]. Regulatory requirements prioritize interpretability, and clinical workflows require it, which restricts the use of black-box models unless their decision-making can be explained[169,173,181]. Mandates on validation, data privacy, and security ensure reliability but potentially slow innovation[171,174]. Regarding the use of AI for closed-loop control, traditional approval pathways struggle with real-time adaptation, as regulations are designed for fixed algorithms rather than dynamic models. While recent frameworks, including the EU AI Act and MHRA's AI Airlock initiative, aim to introduce adaptive oversight mechanisms, striking a balance between innovation and accountability remains a key bottleneck in AI-driven neurotechnologies[181,182].

A final barrier to translation is the heterogeneity that pervades much of pre-clinical, clinical and non-clinical research into neuromodulation. Variability in study protocols and methodologies heavily influences how neuromodulation devices are designed, packaged, and implemented. In clinical trials, methodological variability paired with small sample sizes severely limits the external validity of findings and thus the potential for generating meta-analytic evidence necessary for clinical uptake. Additionally, inaccuracies in clinical diagnosis, undetected comorbidities, and the inherent heterogeneity among clinical populations can distort both the selection of biomarkers and the corresponding stimulation paradigms. This weakens the causal link between symptom severity and modulation outcomes[183]. Small studies with divergent designs also yield inconsistent data formats and variable data quality, complicating the creation of robust, generalizable machine learning algorithms[184]. Models often become overfitted when trained on limited or highly variable datasets, hindering their ability to perform across broader populations. The extensive parameter space in closed-loop systems, ranging from different neural biomarkers to a multitude of stimulation parameters, makes it difficult to compare or replicate findings across studies[185–188]. To overcome these issues, researchers must develop standardized datasets that reflect diverse patient traits and conditions[189,190]. Validating these approaches requires rigorous effect size evaluations, adequate sample sizes, and clear clinical relevance, which are benchmarks that can be difficult to achieve without consistent methodologies and regulatory alignment.

In summary, even the most advanced neuromodulation technology will face adoption barriers if it does not align with regulatory requirements or the needs of stakeholders. Both of these, in addition to large, robust and methodologically consistent human trials, are necessary for the deployment of neurotechnology systems for clinical and non-clinical applications. Engaging stakeholders throughout development, incorporating feedback early, and refining verification and validation processes post-release are crucial for ensuring successful implementation and long-term usability.

*Resolving this trade-off, developers must embed usability, interpretability, and stakeholder engagement into every stage of design, ensuring alignment with regulatory and translation workflows.*

## 4. Insights gathered from discussion of trade-offs

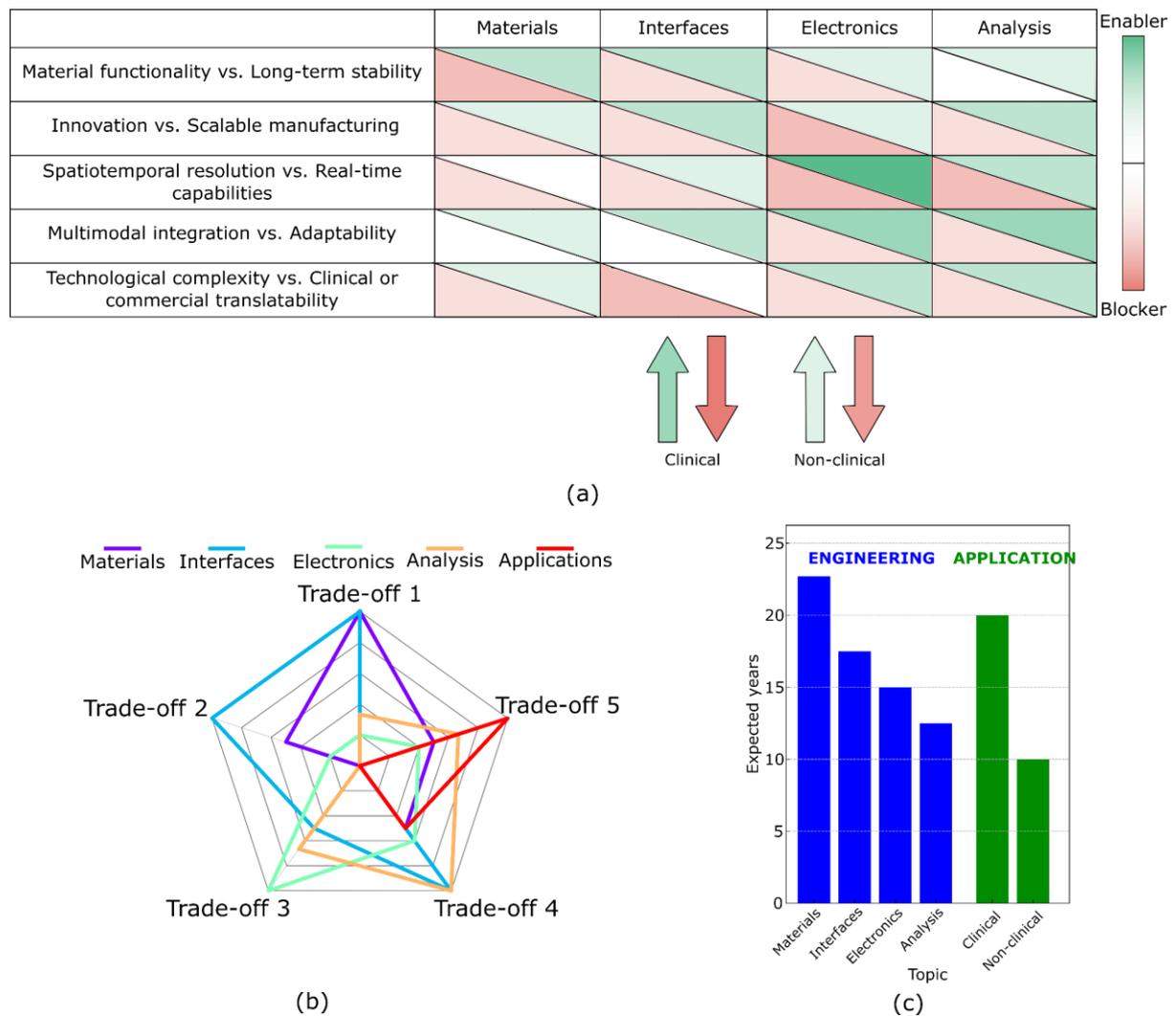

**Figure 2. Multidimensional analysis of technical trade-offs and collaborative strategies in neurotechnology development.** **(a)** Influence matrix illustrating how four technical domains (Materials, Interfaces, Electronics and Analysis) facilitate or constrain progress across five key trade-offs (1. material functionality vs. long-term stability, 2. innovation vs. scalable manufacturing, 3. spatiotemporal resolution vs. real-time capabilities, 4. multimodal integration vs. adaptability and technological complexity vs. clinical or commercial translatability) in neurotechnology. Clinical and non-clinical applications serve as overarching drivers, exerting differing levels of influence across all the trade-offs, as depicted by the intensity of the color. **(b)** Collaborative opportunity map highlighting domain-level interactions required to resolve each trade-off. **(c)** Estimated timelines for resolving each trade-off, based on a consensus discussion during and after the ECRs workshop.

Understanding the role of each technical domain in shaping key trade-offs is crucial for advancing open-loop and closed-loop neurotechnologies. Figure 2a presents an influence matrix developed by ECRs that maps how each domain can either facilitate or constrain progress, helping to identify targeted strategies to improve both reliability and efficiency. The matrix reveals that Materials and Interfaces exert a strong influence on the first two trade-offs: material functionality vs. long-term stability (Trade-off 1) and innovation vs. scalable manufacturing (Trade-off 2), highlighting their critical role in ensuring reliability over time. In contrast, Electronics and Analysis are more strongly associated with the third and fourth trade-offs: spatiotemporal resolution vs. real-time capabilities (Trade-off 3) and multiscale multimodal integration vs. adaptability (Trade-off 4), reflecting their importance in advancing the efficiency and responsiveness of neurotechnologies. For the fifth trade-off: technological complexity vs. clinical or commercial translatability (Trade-off 5), all four technical domains show relatively balanced influence, underscoring that overcoming translational barriers requires a coordinated effort across disciplines. Finally, clinical and non-clinical applications sit above the technical matrix as overarching drivers, shaping priorities and defining success metrics for the development process as a whole.

While Figure 2a provides a snapshot of which domains support or hinder progress, it does not capture the complexity of multidisciplinary collaboration needed to resolve trade-offs. Figure 2b addresses this gap by identifying collaborative opportunities within and across domains. For Trade-off 1, achieving both material functionality and long-term stability requires tight coordination between Materials and Interfaces: material scientists must understand the available interfaces and front-end fabrication processes in order to design materials that are compatible with existing integration methods. Conversely, manufacturing engineers should be familiar with emerging materials and adapt their designs and interface structures accordingly, expanding the range of feasible material properties that can be successfully integrated. Similarly, the advancement of innovative development and scalable manufacturing (Trade-off 2) relies heavily on Interfaces, while still requiring critical input from the Materials domain. Addressing the trade-off between spatiotemporal resolution and real-time capabilities (Trade-off 3) necessitates the coordinated efforts of Interfaces, Electronics, and Analysis. Each domain contributes distinct expertise and resources, and effective trade-off resolution depends on their ability to compensate for each other's limitations. Multi-modal integration and data management (Trade-off 4) primarily involve Interfaces and Analysis, with Electronics also playing a supporting role. Signal processing and computing engineers must develop novel algorithms that streamline integration tasks for manufacturing engineers, while hardware can be optimized to deliver more robust and structured data for downstream analysis. Finally, resolving the trade-off between technological complexity and clinical or commercial translatability requires close interaction between clinical and non-clinical users and the software engineers responsible for the user interfaces in neurotechnologies. User-facing electronic components must be well-documented and accessible, while materials used in these systems must be safe, reliable, and durable.

Figure 2c illustrates the expected timeline for resolving each fundamental trade-off, as estimated by early-career researchers across all domains. A clear trend emerges: the anticipated resolution timeline tends to shorten as the level of physical interaction decreases across the technical domains. This is expected since interacting with the real world introduces an additional level of complexity and makes interactive feedback a slower process. The longest estimated timeline is associated with Materials, projected at approximately 20–25 years. Subsequent topics reduce their estimated time by approximately 5 years. In contrast, timelines in the application domains vary significantly between clinical (20 years) and non-clinical (10 years) contexts. This disparity is expected, given the slower pace of clinical neurotechnology development, largely due to stringent regulatory requirements. Non-clinical applications are expected to overcome key bottlenecks more quickly, benefiting from less burdensome regulation and stronger commercial incentives. According to the surveyed researchers, non-clinical applications will likely be ready for deployment as soon as the relevant technical trade-offs are resolved, an ideal alignment. However, clinical applications must proactively address development delays. Strategies should be considered to accelerate the readiness of the clinical use space, ensuring preparation for adoption once the underlying technologies are mature.

## 5. The future of neurotechnologies

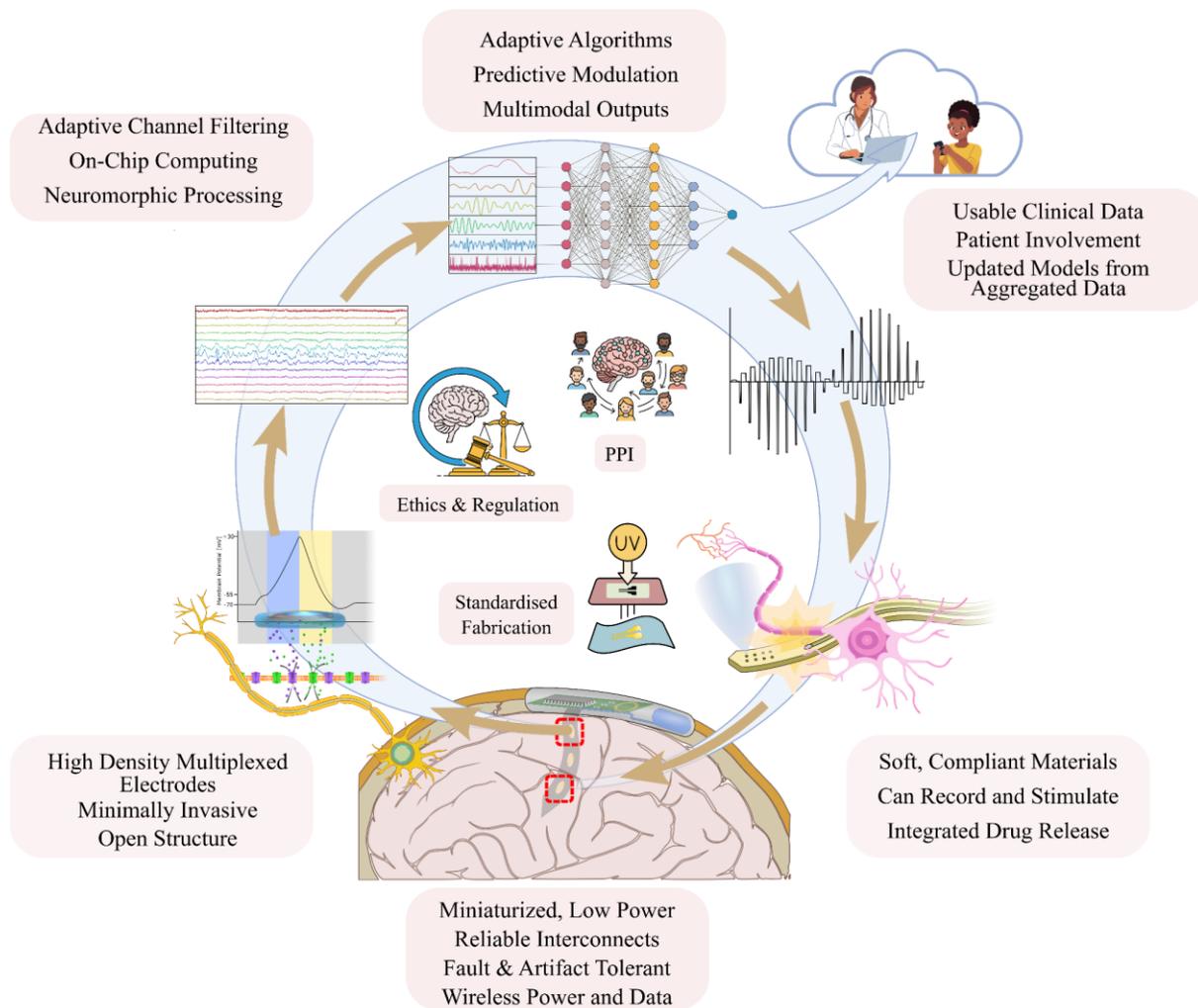

**Figure 3. A vision of the future for closed-loop neuromodulation.** This schematic depicts the journey of neural signals from detection of single neuron depolarization events (spikes), through recording of many-site aggregated signals (local field potentials), and the deconvolution of these into discrete brain waves, on to the processing and clinical decision making, and finally the therapeutic neuromodulation intervention. At the core of this progress lie ethical and regulatory considerations, patient and public involvement (PPI), and the need for standardized fabrication processes. Each step has many possible futures, the text reflects our collective vision of what an adaptive neuromodulation system should comprise.

Closed-loop neurotechnology systems have the potential to revolutionize the clinical treatment of chronic disorders and, in so doing, significantly relieve the burden of global diseases[191]. Our collective vision on the future of adaptive neuromodulation is depicted in Figure 3. Within the next 5 years, we can expect a series of transformative successes in correcting localized anatomical deficits where the intended effect is known and the clinical benefit overt (e.g. spinal cord bypasses[79,192], neurally-integrated prostheses[193] and advanced speech decoding[194]), enabled by early resolution of trade-offs in electronics, interfaces, and signal analysis. Within the same timeframe, we can anticipate the effective application of closed-loop systems to brain disorders characterized by altered connectivity and network dynamics[195], aided by advanced and individualized models of the brain[196,197], as well as multi-modal[198] and multi-site approaches[199] as multimodal integration and adaptability improve (Trade-off4). Their full scope of application, however, is likely to be far broader. Within the next 10 years, it will encompass conditions beyond the nervous system such as cancer[200], functional gastrointestinal disorders[201], diabetes[202] and overactive bladder[203], as scalable manufacturing and real-time capabilities mature (Trade-offs 2 and 3). Powerful models, comparable to today's large language models[204], will emerge from vast amounts of anonymized data, uploaded from individual devices to a shared cloud database[205], and will play a key role in supporting clinical decisions at scale[191,197].

In 5 to 10 years, we can expect the commercial neurotechnology sector to refine technologies for data acquisition[206–209], implantation[210], and brain-computer interfacing to a point where communication is seamless, decision-making, memory, and attention are easily enhanced[211], and productivity is high across sectors[212], driven by progress in usability and non-clinical translatability (Trade-off 5). The non-clinical use of adaptive systems to enhance desirable neural states will proliferate horizontally into sectors such as education (for learning[213]), sports (for reaction time and muscular enhancement[214]), entertainment (for responsive virtual environments[215] and biologically-attuned virtual avatars[216]), and the military (for enhanced physical and mental capabilities[217,218,219]). At this point, the ubiquitous generation of 'neurodata' may well incentivize external parties, such as employers, marketing teams, or governments, to tailor their strategies for management, advertising, or policy in order to leverage insights from neural data and optimally influence human behavior[214]. Questions as to the legality of data sharing and access will be of utmost relevance[220] and a balanced approach to regulation will be essential[221].

As systems advance further, we may see the incorporation of an 'external loop' designed to monitor the user's external environment in real time[222]. This will enable context-specific modulations, such as pre-emptive calming in the event of PTSD triggers, or a sense of reward when presented with healthy food options, to both treat disorder and enhance desirable behavior, respectively. In several decades' time, after years of hardware and software refinement, once long-term stability and regulatory integration are achieved, precision electronic medicine (control down to the individual neuron) may be achieved[223]. In this event, no disorder will be too complex to be treated. Indeed, treatment will simply be a case of temporarily 'rehabilitating' aberrant nervous systems to within normative bounds[205], or of altering a neurodevelopmental trajectory to prevent the onset of illness entirely[224]. At a societal level, the ubiquitous enablement of super-human cognition would allow for unprecedented levels of productivity and, arguably, the ushering in of a technological singularity[225].

## 5.1. Preparing for the ethical landscape of tomorrow's neurotechnology

Neurotechnologies raise urgent ethical and regulatory challenges that must be addressed proactively, not retroactively, as they move from research to real-world use. Ethical frameworks must address issues of ongoing consent, longitudinal feedback effects, and shared liability among clinicians, manufacturers, and developers.  In addition, closed-loop systems that autonomously adapt to physiological and behavioral states in real time introduce new risks beyond traditional neurotechnologies. Their ability to shape neural activity over time raises concerns about loss of user agency, unintended behavioral effects, and long-term neuroplastic consequences. Because adaptive systems dynamically evolve post-deployment, traditional models of one-time informed consent and fixed accountability are insufficient. Additionally, users and clinicians must be able to understand, audit, and intervene in system behavior over time.

These require multidisciplinary collaboration between ethicists, regulators, clinicians, engineers, and patient groups. Ethical guidelines should evolve alongside flexible and adaptive regulatory frameworks, establishing globally harmonized standards for algorithmic transparency, data security, privacy, and autonomy. Standardized Patient and Public Involvement (PPI) guidelines should be followed by clinicians and researchers from early design stages to ensure long-term bidirectional trust, equity, and relevance, and minimize the risk of user abandonment due to discontinuation of support[226]. PPI participants should have a say in shaping guidelines regarding the ethical use of these technologies.

Expanding on this idea, development pipelines for non-clinical technologies, particularly those aimed at wellness or augmentation, should also be informed by rigorous "gold-standard" frameworks co-developed by regulators, scientists, and industry. While such technologies may be commercially driven or speculative in nature, transparency and relevance to the users' benefits must remain central to their reporting standards. Here, collaboration between academic researchers and industry partners is essential, not only for generating robust empirical data, but also for creating an evidence base that supports broader adoption. This is particularly critical in the wellness and consumer neurotechnology sectors, where devices often escape medical oversight and are seldom empirically validated.

However, the current landscape of neurotechnology regulation reflects a tension between the need for robust assurances regarding safety, efficacy, and data privacy, and the urgency of making these technologies accessible and inclusive to clinicians, patients, and consumers[227]. The lengthy and costly pathways for commercialization, regulatory approval, and ethical oversight often deter industry engagement, hindering the development of scalable and impactful neurotechnologies. A balance must be struck: enabling innovation while safeguarding users[228].

While existing programs like the Brain Research Through Advancing Innovative Neurotechnologies (BRAIN) [webBRAIN], Advanced Research + Invention Agency (ARIA) [Precision Neurotechnologies], and Horizon Europe [horizonWeb] broadly support neuroscientific research, dedicated initiatives focusing specifically on closed-loop neuromodulation are still needed. Recent efforts, such as the UK MHRA's AIaMD Change Programme, the EU's Artificial Intelligence Act and MDR, and the FDA's Breakthrough Devices and Precertification programs, represent important steps toward regulating adaptive technologies. In June 2025, the UK introduced a major update to its medical device regulations, streamlining approval pathways and strengthening post-market oversight, an important step toward aligning safety standards with the pace of neurotechnology innovation[229]. Moreover, in April 2025 the UN Human Rights Council adopted a resolution endorsing human rights approach to neurotechnologies[230]. However, closed-loop neurotechnologies again introduce unique safety, accountability, and long-term use challenges. These features fall outside the scope of most existing frameworks. While technical standards such as IEC 60601-1-10 (covering physiologic adaptive controllers) and ISO 14971 (risk management for medical devices) provide important guidance, their adoption remains inconsistent, especially in non-traditional or early-stage applications. Addressing these challenges will require more targeted, internationally harmonized regulatory approaches that integrate standards for algorithmic transparency, continuous performance validation, and shared liability, for both clinical and non-clinical applications.

Finally, policies must anticipate societal consequences such as cognitive surveillance, inequitable access, and shifts in human autonomy[228]. By aligning ethical responsibility with innovation pathways, we can support the safe and meaningful integration of open- and closed-loop neurotechnologies into clinical and everyday life.

**5.2. Call to action for stakeholders**

The promise of neurotechnologies will only be realized through coordinated, cross-sectoral action. We believe:

1. Researchers and clinicians must adopt and refine robust, standardized protocols for both clinical and preclinical studies enabling reproducibility, meta-analysis, and translational insight. At the same time, fostering open-source development is essential to accelerate collaboration, promote transparency, and enable rapid adaptation across research and clinical settings.

2. Regulators and journal editors should coordinate efforts to enforce transparent reporting standards while safeguarding innovation balancing accountability with flexibility in emerging fields.

3. Industry partners, including those in the wellness and consumer neurotechnology sectors, must commit to sharing efficacy and safety data openly, especially for interventions that fall outside formal regulatory oversight. Wherever possible, companies should adopt open-source practices to promote transparency, accelerate innovation, and enable broader scientific scrutiny. Open-source development ensures that essential knowledge is not confined within the boundaries of individual commercial interests, whose priorities and stability may shift over time. This approach also facilitates interoperability, reproducibility, and collaboration across academic, clinical, and industrial domains.

4. Governments must invest in the manufacturing landscape at national and international levels to enable the translation of promising neurotechnologies from research to clinical implementation.

5. PPI should be embedded from the earliest stages of design, not as an afterthought, to ensure real-world relevance, equity, and public trust in neurotechnology deployment.

6. Funders must launch targeted initiatives to de-risk the critical bottleneck between innovation and scalable manufacturing, a transition that currently stalls many promising neurotechnologies in preclinical stages. These initiatives should go beyond general neuroscience or medical device innovation, focusing specifically on the unique challenges of systems. Support should include long-term validation across diverse populations and sustained funding mechanisms for technologies deployed in decentralized or low-resource settings, ensuring equitable access and real-world applicability.

7. Investors must also recognize the unique timelines of neurotechnology and commit to "patient capital" models that extend beyond conventional 3–5-year return horizons. Traditional venture capital expectations are misaligned with the reality of clinical pilot trials, regulatory approval, and insurance adoption. Dedicated investment vehicles, hybrid funding models, and public–private partnerships are essential to sustain innovation through the long translational pathway. Without such patient capital, the field risks premature abandonment of promising technologies before they can reach patients.

8. Policymakers must establish clear mandates for long-term oversight, data governance, and ethical accountability, especially in cases where responsibility becomes diffused after initial funding or deployment. Policies must enable innovation while ensuring equitable access, protecting autonomy, and preventing misuse particularly in sensitive domains like cognitive modulation.

9. Finally, mechanisms for post-funding responsibility and longitudinal oversight must be clarified. Long-term monitoring of systems often falls into institutional grey zones, risking both user safety and innovation stagnation. Sustainable frameworks are needed to ensure that responsibility for safety, efficacy, and equitable access persists beyond initial trials or grants.

We urge all stakeholders to move beyond siloed development toward a shared roadmap for the design, testing, and deployment of neuromodulation systems. In doing so, we can accelerate the creation of equitable, effective, and responsive technologies that adapt to real-time physiology, fundamentally reshaping patient care across a range of disorders. The time to act is now: without coordinated effort, innovation risks outpacing both evidence and impact.

**Final notes**

This roadmap reflects the collective insights of a specific group of early-career researchers who participated in a focused workshop at a particular point in time. While it captures a broad and interdisciplinary perspective, it does not represent an exhaustive or universally agreed-upon vision. The field of neurotechnology is evolving rapidly, and emerging discoveries, technologies, and societal needs may shift priorities in unforeseen ways. As such, this roadmap should be viewed as a living document, one that invites ongoing dialogue, revision, and expansion as the field progresses.

**Acknowledgements**

We thank Prof. Rylie Green, Prof. George Malliaras, Prof. Emm Drakakis, Prof. Marcus Kaiser, Prof. Antonio Valentin, Prof. Tamar Makin, Prof. Andrew Jackson, and for their valuable contributions as speakers at the Cambridge Closed-Loop Neurotechnology Workshop in November 2024, and for their thoughtful feedback on the manuscript. Their insights and suggestions were instrumental in shaping and refining this work. We also gratefully acknowledge the support of the CloseNIT network for funding the workshop and enabling the collaboration that led to this paper. We thank the contribution of Dr Nic Shackle, Dr Sepehr Shirani, Dr Jichun Li, and Eduardo Marques for their insightful discussions during the Workshop. AG acknowledges support from the Royal Commission for the Exhibition of 1851 Research Fellowship, Royal Academy for the Engineering Research Fellowship (#RF-2324-23-284) and Rosetrees Research Fellowship. JGT, MM, and LA acknowledge the ongoing support of Dr Christopher Proctor. NM acknowledges the ongoing support of Prof. Timothy Constandinou. NT acknowledges the ongoing support from PI's Prof. David Holder and Dr Kirill Aristovich and funding from Medical Research Council Grant MR/Z504555/1. SW is supported by UK Research and Innovation [UKRI Centre for Doctoral Training in AI for Healthcare grant number EP/S023283/1]. ISL is supported by EPSRC Neurotechnology for Chronic Pain Network grant EPSRC/MRC EP/W03509X/1. EC acknowledges the ongoing support of Prof. Rylie Green. This work was funded by UKRI grants. For the purpose of open access, the author has applied a Creative Commons Attribution (CC BY) licence to any Author Accepted Manuscript version arising.